\documentclass[aps,reprint,amsmath, amssymb,groupedaddress,floatfix]{revtex4-1}
\usepackage{natbib}
\usepackage{graphicx}
\usepackage{bm}
\usepackage{color,soul}
\usepackage{hyperref}
\usepackage{float}
\usepackage{csquotes}
\usepackage{color,soul}
\usepackage{hyperref}
\hypersetup{colorlinks=true, urlcolor=blue, citecolor=blue}
\usepackage{longtable}
\usepackage{tabularx}

\newcommand{\cgb}{{\rm CsGeBr$_3$}}

\begin{document}

\title{Coexistence of polar and antipolar phases in ferroelectric halide perovskite CsGeBr$_3$}

\author{Ravi Kashikar}
\email{ravik@usf.edu}
\affiliation{Department of Physics, University of South Florida, Tampa, Florida 33620, USA}
\author{S. Lisenkov}
\author{I. Ponomareva}
\email{iponomar@usf.edu}
\affiliation{Department of Physics, University of South Florida, Tampa, Florida 33620, USA}

\date{\today}

\begin{abstract}
Recently ferroelectricity has been demonstrated in the family of halide perovskites: CsGeX$_3$ (X=Cl, Br, I). We develop a first-principles-based computational approach to probe finite-temperature properties of \cgb. Our computations reveal the existence of a dynamic antipolar phase above the Curie temperature. The phase can be stabilized below the Curie temperature through quenching. Furthermore, competition between the polar and antipolar phases results in the formation of rare polar-antipolar domains.  We demonstrate that such polar-antipolar domains can result in the formation of unusual polarization fields with head-to-head and tail-to-tail dipole arrangements, which not only offers an explanation for the recent experimental findings but could also lead to the discovery of novel topological polar structures.

\end{abstract}
\maketitle

\section{Introduction}
Inorganic halide perovskites of the form ABX$_3$, where A=Cs or Rb; B=Ge, Sn, Pb and X=Cl, Br, I  have gained significant attention because of their  optoelectronic properties\cite{Intro_1, Intro_2, Intro_3}.   These perovskites exhibit direct bandgap, low thermal conductivity, high absorption coefficient,  high carrier mobility, and long carrier diffusion length\cite{Intro_4}. These properties find application in solar cells, light-emitting diodes, semiconductor lasers, photocatalysts, and memristors. These properties are further tunable with composition alterations and external fields, such as  pressure, uniaxial strain, and biaxial strain\cite{compo, strain_1, strain_2}.   The general structural arrangement consists of BX$_6$ octahedral cages surrounded by A site entity. Many halide perovskites undergo phase transitions under pressure and/or temperature to lower symmetry phases\cite{alaei2021polymorphism, phase}. Among these, Sn and Pb-based crystal phases are centrosymmetric in nature and have multiple phase transitions with temperature\cite{phase}.  For example, CsSnX$_3$ and CsPbX$_3$ exhibit cubic to tetragonal and  tetragonal to orthorhombic phase transitions due to octahedral rotations because their Goldschmidt tolerance factor  is less than 1. For CsGeBr$_3$ and CsGeCl$_3$, the tolerance factor  is greater than 1, which favors  the B site displacements and leads to stabilization of the polar phase. Ge-based perovskites have been proposed to exhibit  order-disorder single-phase transition from  cubic $Pm\bar{3}m$ phase to  polar $R3m$  phase  at 428~K, 511~K, and 550~K for X=Cl, Br, and I, respectively\cite{thiele1987kristallstrukturen}. Recent experiments have demonstrated  ferroelectricity in CsGeX$_3$ (X = Cl, Br, I) and reported spontaneous polarization of 12-15~$\mu$C/cm$^2$ for X=Br, 20~$\mu$C/cm$^2$ for X=I\cite{CGX_Ferro}. So far, there are no experimental polarization reports for X=Cl. With this, CsGeX$_3$ are the only inorganic halide perovskites where the ferroelectricity has been demonstrated experimentally.  Furthermore, the bandgap of these materials lies between 1.6 to 3.3~eV and is suitable for ferroelectric semiconducting devices\cite{CGX_Ferro}.

At the moment, the understanding of the ferroelectric phase transitions in these materials is lacking. How do the order parameters emerge and evolve with temperature? What is the character of the phase transition and its order? How does the material's response to the electric field depend on the temperature? Does the phase transition in this family have some distinct features? Are there any hidden phases or unusual domains/structures? Do the phase transitions have unique dynamical signatures? Atomistic first-principles-based simulations, such as effective  Hamiltonian\cite{Effect_H0, Effective_H1}, is an 
excellent tool to probe the atomistic nature of the phase transitions in complex ferroics and answer some of the aforementioned fundamental questions. Such methodology has previously been used to study phase transitions in ferroelectrics\cite{Effect_H0, PTO_phase}, ferroelectric alloys\cite{BST_1, PZT_1, BZT_1, BZT}, multiferroics\cite{BFO}, antiferroelectrics\cite{PZo_phase} and predict novel phases and phenomena such as ferroelectric vortices\cite{Nanoplate,Naumov2004}, nanostripes\cite{stripe}, bubbles\cite{bubble, bubble_2}, ferroelectric phases in antiferroelectrics\cite{AntiF_Ferro}, and electromagnons\cite{magnon} to name a few. 

In this work, we aim to: (i)  develop an effective Hamiltonian  approach to  study finite temperature properties of \cgb; (ii)  utilize this approach to probe ferroelectricity and associated phase transition in this material in order to answer some of the fundamental questions raised above; (iii) to predict the existence of both polar and antipolar phases in \cgb; (iv) to reveal the possibility of these  phase coexistence leading to the formation of polar-antipolar domains, which result in unusual polarization fields. 

\begin{figure*}
\centering
\includegraphics[width=0.8\textwidth]{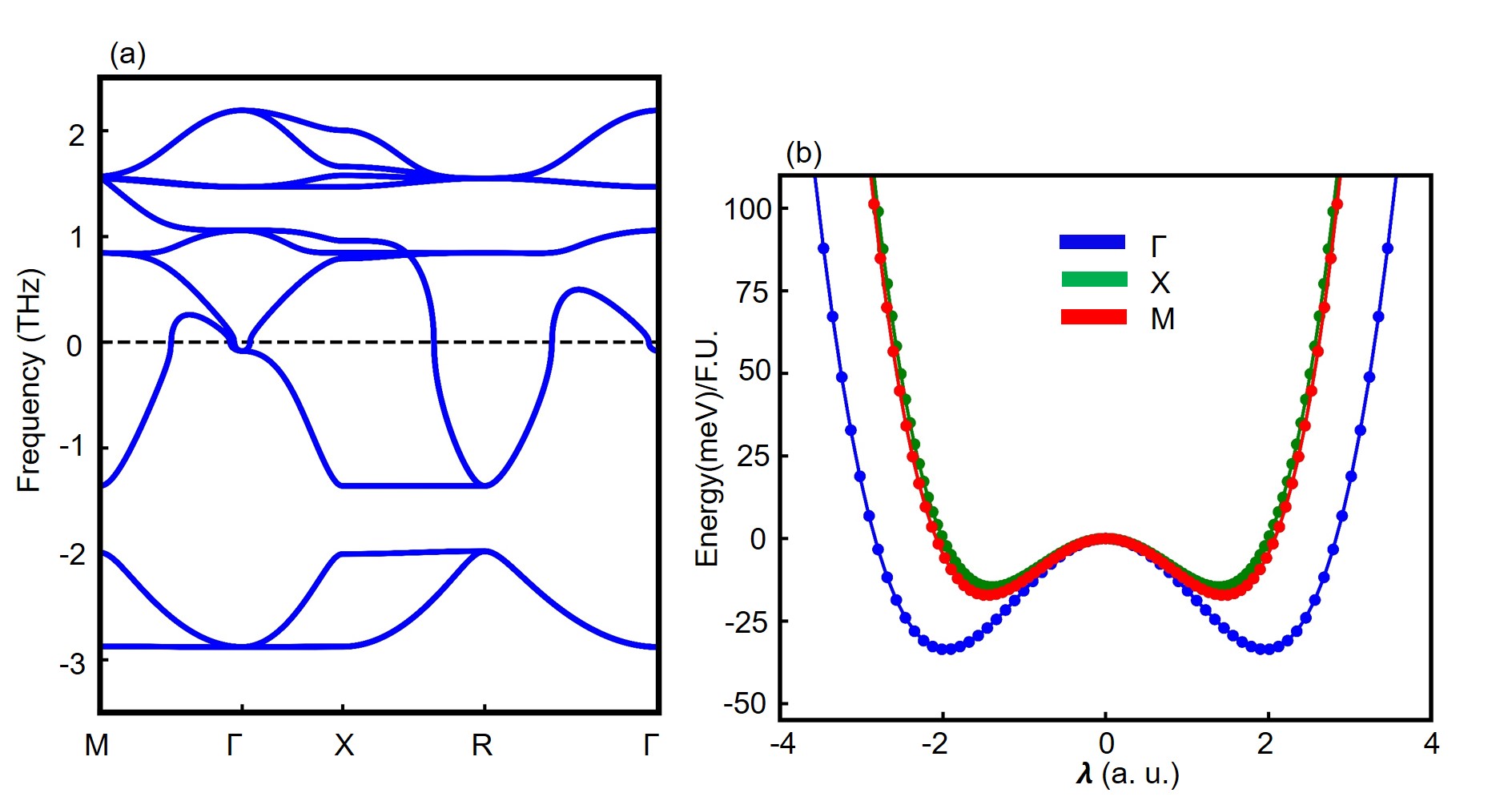}
\caption{(a) Phonon dispersion of CsGeBr$_3$. (b) Energy along the directions of the CsGeBr$_3$  unstable modes at  $\Gamma$, M and X. }
\label{fig1}
\end{figure*}

We begin by investigating phonon instabilities in the cubic phase of \cgb\, using density functional theory (DFT)  computations as implemented in $\textsc{VASP}$ package \cite{vasp1,vasp2}.  Technically, we use  projector-augmented basis set (PAW) \cite{paw} within GGA-PBE  \cite{PhysRevLett.77.3865} approximation.  Plane wave energy cutoff of 550~eV and k-point grid density 0.2~$\AA^{-1}$ were used for all our ground state calculations. However, for computing phonon dispersions, elastic constants and macroscopic dielectric tensor higher energy cutoff (800~eV) and denser k-point mesh ($16\times16\times16$) were used to ensure convergence of the above-mentioned properties.  Figure~\ref{fig1}(a) reports the phonon dispersion  for cubic $Pm\Bar{3}m$ phase of \cgb\, computed using density functional perturbation theory (DFPT) and post-processed using $\textsc{PHONOPY}$ package \cite{phonopy}. The phonon dispersion predicts the existence of an unstable branch with the strongest instabilities in $\Gamma$, $X$, and $M$ points. This is in agreement with previous computational findings\cite{phase}.

Note, that we have also used the LDA exchange-correlation functional \cite{PhysRevLett.45.566}. However, we found that it predicts the cubic phase to be more stable than $R3m$ (which contradicts to experiment), and computed phonon dispersion within LDA approximation does not posses unstable modes. Therefore, we use the PBE approach for our methodological developments. Next, we compute the energy profile along the eigenvector direction of the unstable phonons at $\Gamma$, $X$, and $M$ points. The associated energy landscape is given in Fig.~\ref{fig1}(b) and suggests the existence of metastable phases in this material, that is, the phases of local minima at the energy surface. The structures that correspond to the minima for $\Gamma$, $X$, and $M$ point instabilities  in Fig.~\ref{fig1} are subjected to full structural relaxation, which resulted in the following phases - $R3m$ (R), $Pmma$ (O) and $P4/nmm$ (T) with the energies with respect to the cubic one of  -86.1, -29.7, and -25.9 meV/f.u., respectively. These structures are associated with antipolar arrangements Ge and Br ions and are provided in Ref. \cite{ourgithub}.  

Next, we use DFT calculations to parameterize the effective Hamiltonian model of Ref.\cite{Effective_H1}. The degrees of freedom for the effective Hamiltonian are local modes (which are proportional to the local dipole moment of the unit cell), homogeneous, and inhomogeneous strains. We use  the eigenvector of unstable mode at $\Gamma$ point to parameterize our Hamiltonian. The vector is dominated by Ge displacement, so we center local modes on  Ge (see Table~\ref{T1}).  The Hamiltonian includes the following interactions: local mode self
energy (harmonic and anharmonic contributions), a long-range
dipole-dipole interaction, a short-range interaction between
local modes, elastic energy, and the interaction between
the local modes and strains. Effective Hamiltonian can be utilized in the framework of Monte Carlo, Molecular Dynamics (MD), and other techniques to simulate finite-temperature properties and has a long  success history \cite{Effective_H1, PTO_phase, PZo_phase,Naumov2004, Vortex,AntiF_Ferro,BFO,Flexo}. 

The parameters of the effective Hamiltonian that we derived for \cgb\  are given in Table~\ref{T1} and can be used with the publicly available software $\textsc{FERAM}$ \cite{PhysRevB.78.104104,feram}.  We note that the set of parameters reproduces very well the lowest energy phonon branch shown in Fig.~\ref{fig1}(a). In particular, it captures competing instabilities in $\Gamma$, $X$, and $M$ points and their relative strength. Table~\ref{T2} compares predictions of the effective Hamiltonian with DFT and some available experimental data and demonstrates its  accuracy, including in capturing the dynamics of the soft mode. 

\begin{table*}
\caption{\label{tab:params}%
First-principles-based parameters for CsGeBr$_3$ in atomic units using the notations of Ref.\cite{Effective_H1} derived from PBE/r$^2$SCAN functionals. }
\begin{ruledtabular}
\begin{tabular}{ccccccc}
\textrm{On-site}&
\textrm{$\kappa_2$}&
3.29$\times$10$^{-3}$/2.50$\times$10$^{-3}$&
\textrm{$\alpha$}&
5.61$\times$10$^{-2}$/6.62$\times$10$^{-2}$&
\textrm{$\gamma$}&
-9.01$\times$10$^{-2}$/-10.6$\times$10$^{-2}$\\
\colrule
 & \textit{j$_1$} & -2.56$\times$10$^{-3}$/-3.09$\times$10$^{-3}$  & \textit{j$_2$} & 7.34$\times$10$^{-4}$/10.34$\times$10$^{-4}$ &  &   \\
Intersite & \textit{j$_3$} & 8.96$\times$10$^{-4}$/10.79$\times$10$^{-4}$ & \textit{j$_4$} & -4.62$\times$10$^{-4}$/-5.62$\times$10$^{-4}$ & \textit{j$_5$} & 4.67$\times$10$^{-4}$/8.19$\times$10$^{-4}$  \\
 & \textit{j$_6$} & 2.12$\times$10$^{-4}$/2.58$\times$10$^{-4}$ & \textit{j$_7$} & 0.00/0.00 &  &   \\
Elastic & \textit{B$_{11}$} & 1.91/2.15 & \textit{B$_{12}$} & 0.42/0.45 & \textit{B$_{44}$} & 0.39/0.46  \\
Coupling & \textit{B$_{1xx}$} & -0.48/-0.51 & \textit{B$_{1yy}$} & -0.16/-0.19  & \textit{B$_{5xz}$} &  -0.01/-0.01 \\
Other & \textit{Z$^{*}$} & 7.69/8.25 & \textit{$\epsilon_{\infty}$} & 7.11/7.00 & \textit{m} & 53.52/48.31 \\
      & \textit{a} ($\AA$) & 5.61/5.54 & & & & \\
Eigenvector & \multicolumn{6}{c}{(-0.01,-0.75,0.66,0.05,0.05)/(-0.01,-0.75,0.66,0.03,0.03)} \\
\end{tabular}
\end{ruledtabular}
\label{T1}
\end{table*}

\begin{table*}
\caption{\label{tab:properties}%
Some ground state properties of \cgb\, as obtained
from DFT, effective Hamiltonian (Heff), and experiment (Exp).
The experimental data are reported for room temperature and taken
from \cite{CGX_Ferro}. The energies are reported with respect to the cubic phase, while ${^\star}$ indicates the estimate for the energy difference between the two phases from the Curie point.  }
\begin{ruledtabular}
\begin{tabular}{cccccccc}
\textrm{} &
\textrm{\textit{$\Delta$E$_T$}}&
\textrm{\textit{$\Delta$E$_O$}}&
\textrm{\textit{$\Delta$E$_R$}}&
\textrm{\textit{a}}&
\textrm{\textit{P$_S$}}&
\textrm{\textit{$\omega_{E}$}} & 
\textrm{\textit{$\omega_{A_1}$}} \\
\textrm{} &
\textrm{(K)}&
\textrm{(K)}&
\textrm{(K)}&
\textrm{(\AA)}&
\textrm{($\mu$C/cm$^2$)}&
\textrm{(cm$^{-1}$)} & 
\textrm{(cm$^{-1}$)} \\

\colrule
DFT & -301 & -345 & -999 & 5.75 & 19.9 & 130.4 & 160.8\\
Heff & -242 & -285  & -862 & 5.72 & 23.4 & 141.5 & 160.8\\
Exp. & - &  - & -511$^*$ & 5.63 & 12-15 (RT)  &  138 & 163\\

\end{tabular}
\end{ruledtabular}
\label{T2}
\end{table*}

\begin{figure*}
\centering
\includegraphics[width=0.8\textwidth]{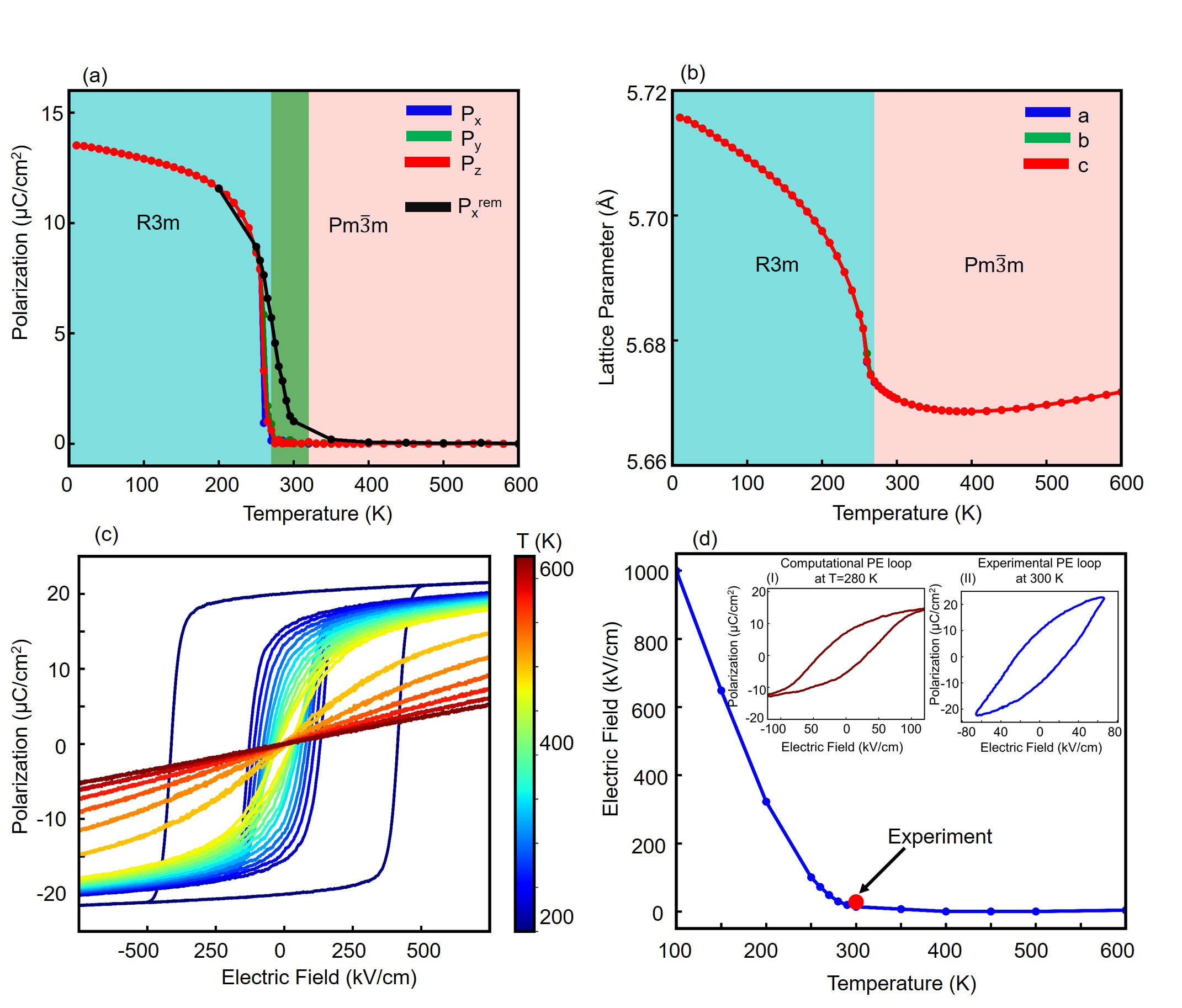}
\caption{ Polarization components (a) and lattice parameters (b) as a function of temperature. In (a) $P_{x,y,z}$ are the Cartesian components of spontaneous polarization computed from simulated annealing. Black data give one Cartesian component of remnant polarization obtained from hysteresis loops. Green shading outlines the region where the polar phase is stabilized by the electric field. Hysteresis loops computed at  1 GHz frequency   in 200~K to 600~K temperature range (c); Coercive field as a function of temperature (d). The inset compares the hysteresis  loops at T=280~K with the RT experimental loop reported in Ref.\cite{CGX_Ferro}. }
\label{fig2}
\end{figure*}

We now use the effective Hamiltonian in the framework of classical  MD simulations to model finite-temperature properties of bulk \cgb. We use a supercell of  30$\times$30$\times$30 unit cells of \cgb\, repeated periodically along three Cartesian directions.  The integration step is 0.5~fs. The Evans-Hoover \cite{rapaport2004art}  thermostat is used to simulate constant temperature. A simulated annealing technique is chosen to obtain equilibrium phases at different temperatures. Technically, we start simulations at 600~K and proceed in steps of 10~K down to 10~K. For each temperature, we performed 500,000 MD steps (0.25~ns) and used half of them for equilibration and half for computing thermal averages.  The polarization is computed as the average dipole moment of the supercell per its volume. Its temperature evolution is given in Fig.~\ref{fig2}(a) and predicts phase transition from nonpolar $Pm\Bar{3}m$ cubic phase to ferroelectric $R3m$ phase at 270~K. The only experimental report to date reports the transition between $Pm\Bar{3}m$ and $R3m$  phase at 511~K on the basis of X-ray diffraction and Raman Spectroscopy \cite{thiele1987kristallstrukturen, CGX_Ferro}. It appears that our effective Hamiltonian underestimates transition temperature, which is  rather common \cite{Effective_H1,PhysRevB.78.104104}. At least part of it could be due to the underestimation of the energy difference between the ground state and cubic phase as compared with the DFT data (see Table \ref{T2}). In fact, the difference of 150~K between the effective Hamiltonian and DFT values could account for the underestimation.  Another contribution could be due to the dependence of the energy prediction on the exchange-correlation functional. Indeed, parametrization with r$^2$SCAN functional \cite{r2scan} resulted in an increase of T$_C$ by 70~K. We notice that the steep onset of the order parameter (polarization) at the Curie point is suggestive of the first-order character of the phase transition.  
\begin{figure*}
\centering
\includegraphics[width=0.75\textwidth]{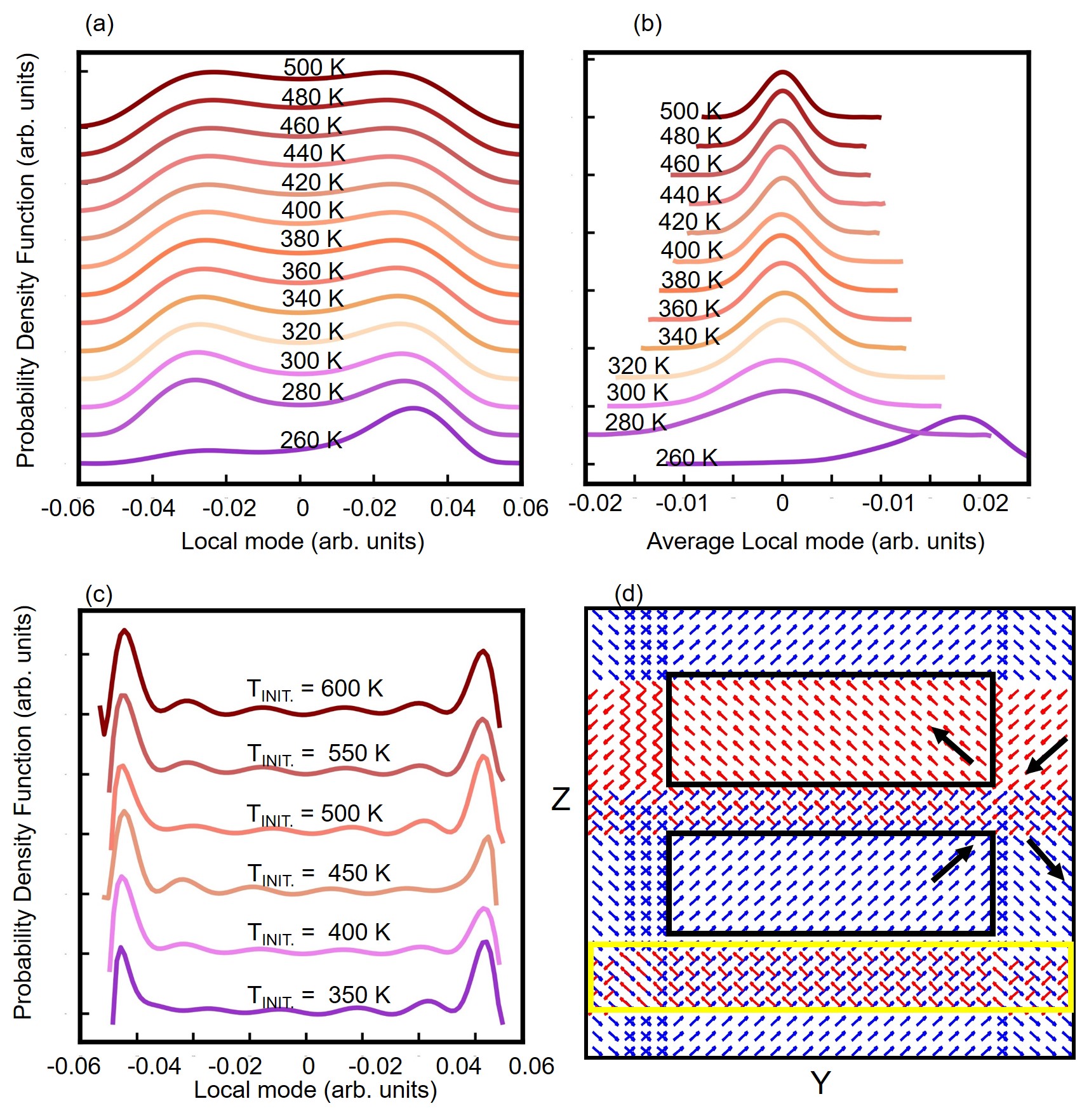}
\caption{Probability density functions for the components of the local modes $u_i$ (a); and for the average components $<u_i>$ (b) computed at different temperatures. Probability density functions for the components of the local modes $u_i$ computed for the quenched structures at 100~K (c). The temperature from which quenching has started is given on the labels. Dipole pattern of \cgb\, at 100~K exhibiting polar (black boxes) and antipolar  (yellow box) domains.  }
\label{fig33}
\end{figure*}
To elucidate the response of polarization to the electric field, we have computed electric hysteresis loops  by applying an ac electric field with frequencies of 1.0, 2.5, 5.0, and 10.0 GHz for all temperatures investigated and show some representative ones in Fig~\ref{fig2}(c). For an ac field frequency of 1 GHz,  we find ferroelectric loops up to 295~K, paraelectric response in the range of 300~K - 400~K, and linear dielectric response for higher temperatures. The loops are used to compute remnant polarization as a function of temperature and added to Fig.~\ref{fig2}(a).    From the comparison between spontaneous and remnant polarization in Fig.~\ref{fig2}(a), we notice that ferroelectricity can be stabilized by the electric field in  the  temperature range of 270-300~K.  The coercive field as a function  of temperature is reported in Fig.~\ref{fig2}(d). We note that the coercive field in computations typically overestimates the experimental one due to the lack of defects in simulated supercells, their finite size, and high-frequency of the applied fields.  Nevertheless, the comparison between the experimental and our computational coercive field is good (see Fig.~\ref{fig2}(d)), which further validates our computational approach. The inset to Fig.~\ref{fig2}(d) shows the comparison between computational and experimental hysteresis loops. 

The temperature evolution of the lattice parameters is given in Fig.~\ref{fig2}(b) and predicts somewhat unusual dependence. Firstly, it does not follow the first-order-like trend exhibited by polarization. The second feature is the dependence on temperature in the $Pm\Bar{3}m$ phase. In our model, the temperature evolution of the lattice parameters originates from the temperature evolution of the local modes owing to the on-site coupling between the local modes and local strains. Therefore, the computational data suggest that the local modes are nonzero even above T$_C$ and even cooperate with each other in some way, which, however, does not result in macroscopic polarization.  To elucidate the origin of this unusual behavior, we compute probability density functions, $\rho(u_i)$, for the Cartesian components of the local modes. Technically, we  ran additional 50,000 MD steps on the equilibrated supercells in the temperature range of 250~K to  600~K and used them to obtain probability density functions. Some representative functions are shown in Fig.~\ref{fig33}(a) and reveal the presence of two maxima in the functions.  Next, we compute the probability density functions for the  average local modes, $\rho(<u_i>)$,  where the average is taken over  50,000 MD steps. 
 The representative functions are shown in Fig.~\ref{fig33}(b) and reveal that the two peaks have  now merged into a single one, which suggests their dynamic nature. 
 We note that this finding also means that experimentally the phase would indeed appear as $Pm\Bar{3}m$. 
  To probe further into the origin of the peaks in $\rho(u_i)$, we quench the equilibrium supercells  that correspond to temperatures from the 350-600~K range down to 100~K. The probability density functions at 100~K  are given in Fig.~\ref{fig33}(a) and exhibit  two well-defined peaks, whose presence proves that quenching stabilizes the dynamical phase that exists above $T_C$.  
  Inspection of the quenched dipole patterns reveals  antipolar domains, which suggests that the dynamic high-temperature phase is the antipolar one. Recall  that  antipolar Pmma(O) and P4/nmm(T) phases are also predicted in DFT computations. Furthermore, the stabilization of antipolar domains as opposed to single-domain phase may suggest that the high-temperature dynamical phase is made up of antipolar nanoregions, the concept similar to the famous polar nanoregions in relaxors \cite{Relaxor_1,Relaxor_2}. Interestingly, such dynamical antipolar domains, or antipolar nanoregions, could explain why the transition in \cgb is believed to be of order-disorder type\cite{thiele1987kristallstrukturen}.

   So far, we learned from simulations   that annealing results in the stabilization of the polar ground state, while ultra-fast quenching yields the antipolar phase with domains. Could the two phases coexist? To answer this question, we carried out quenching with a lower cooling rate; that is, we quenched the supercell from  600~K down to 100~K in steps of 50~K, using 50,000 MD steps for each temperature.  Indeed, in this case, we find the coexistence of polar and antipolar domains as shown in Fig.~\ref{fig33}(d), which to the best of our knowledge, is extremely rare.  The possibility of the coexistence of polar and antipolar domains has been considered in hafnia films\cite{Cheng2022}. Moreover, such coexistence leads to the formation of head-to-head and tail-to-tail domains, as shown in Fig.~\ref{fig33}(d), which are energetically unfavored in traditional ferroelectrics.  Interestingly, such domains are in good agreement with experimentally reported ones from Ref.\cite{CGX_Ferro}. Moreover, the stabilization of such unusual domains could also mean that new topological polar phases could be discovered in the materials with polar/antipolar phase competition.   Experimentally, the stabilization of polar-antipolar domains may be induced by  the thermal history. For example, the samples that have not been annealed or not annealed to high enough temperatures may exhibit polar-antipolar domains similar to the ones predicted in simulations. We note, that all the main findings have been reproduced with parameters obtained with r2SCAN functional. 

   In summary, we developed a first-principles-based computational methodology to investigate the finite-temperature properties of ferroelectric halide perovskite \cgb. The application of the methodology to bulk \cgb\ predicts a single-phase transition from paraelectric cubic $Pm\Bar{3}m$ phase to ferroelectric rhombohedral phase $R3m$ in agreement with experimental observation. However, it also reveals a dynamical antipolar phase that exists above the Curie temperature. Such an antipolar phase can be stabilized through rapid quenching from high temperatures and can compete with the polar phase. The latter competition was found to result in the formation of polar-antipolar domains, which allow for exotic domain configurations that are energetically prohibitive in traditional ferroelectrics. In fact, the only known mechanism to stabilize such domains is through free-charge compensation. The presence of competing antipolar phases offers a different route to such domains and may lead to new forms of topological polar structures. We believe that our work will promote discoveries of novel properties and functionality associated with polar-antipolar phase competitions in halide perovskites.  \\\\\\


This work was supported by the U.S.
Department of Energy, Office of Basic Energy Sciences, Division of Materials Sciences and Engineering under Grant No. DE-SC0005245.  Computational support was provided by the National Energy Research Scientific Computing Center (NERSC), a U.S. Department of Energy, Office of Science User Facility located at Lawrence Berkeley National Laboratory, operated under Contract No. DE-AC02-05CH11231 using NERSC award BES-ERCAP-0025236.



\bibliography{paper}

\end{document}